\documentclass{article}
\usepackage{graphicx} 
\usepackage[T1]{fontenc}
\usepackage{babel}
\usepackage{amsmath}
\usepackage{amsfonts}
\usepackage{amssymb}
\usepackage{amsthm}
\usepackage{amsfonts}
\usepackage{mathtools}
\usepackage{physics}
\usepackage{tikz}
\usepackage{authblk}
\usepackage{dsfont}
\usepackage{hyperref}
\hypersetup{colorlinks,allcolors=black}
\usepackage[a4paper, total={6in, 8in}]{geometry}

\newtheorem{theorem}{Theorem}

\newtheorem{definition}[theorem]{Definition}

\newcommand{\ext}[1]{\underset{#1}{\text{ext}}}
\providecommand{\keywords}[1]
{
  \small	
  \textbf{\textit{Keywords---}} #1
}

\title{Geometric methods in quantum information \\ and entanglement variational principle}
\author[1,3,*]{Daniele Iannotti}
\author[1,2,3]{Alioscia Hamma}
\affil[1]{Scuola Superiore Meridionale\\ Largo S. Marcellino 10, 80138 Napoli, Italy}
\affil[2]{Università degli Studi di Napoli Federico II \\ Dipartimento di Fisica Ettore Pancini\\ Complesso Universitario di Monte S. Angelo ed. 6 \\ via Cintia, 80126, Napoli, Italia}
\affil[3]{Istituto Nazionale di Fisica Nucleare (INFN) Sezione di Napoli\\ Complesso Universitario di Monte S. Angelo ed. 6 \\ via Cintia, 80126, Napoli, Italia}
\affil[*]{(Corresponding author: d.iannotti@ssmeridionale.it)}
\date{}

\begin{document}
\maketitle
\usetikzlibrary{shapes.geometric} 
\usetikzlibrary{shadows.blur}

\begin{abstract}
    Geometrical methods in quantum information are very promising for both providing technical tools and intuition into difficult control or optimization problems. Moreover, they are of fundamental importance in connecting pure geometrical theories, like GR, to quantum mechanics, like in the AdS/CFT correspondence. In this paper, we first make a survey of the most important settings in which geometrical methods have proven useful to quantum information theory. Then, we lay down a general framework for an action principle for quantum resources like entanglement, coherence, and anti-flatness. We  discuss the case of a two-qubit system. 
\end{abstract}

\newpage
\section{Introduction}
A crucial advantage of physics lies in the fact that is possible, up to some non-trivial complications, to map a problem into another where one has a better understanding of both mathematics and physics.
Examples span from the standard method of image charges to more complicated scenarios, e.g. the mapping between the continuum limit of the XY model at low temperatures and a two-dimensional Coulomb gas.
In the former case, to calculate or visualize the distribution of the electric field of a charge in the proximity of a conducting surface one adds a fictitious opposite charge across the surface\footnote{The method lies on certain validity conditions, i.e. a corollary of the uniqueness theorem for Poisson's equation. Considering certain method validity will be crucial for our article in the following chapters.}. 
This mathematical trick does not state any deeper meaning of the fictitious charge except a simpler way to compute the desired quantity, no more than what one does with ghost particles in quantum field theory (QFT).

Geometrical methods in theoretical physics have a long-lasting tradition \cite{nakahara2018geometry} in this realm and they have provided profound insights into reimagining conventional physics through the lens of geometric quantities. 
This approach serves as a powerful tool for creating a lucid conceptual framework.

Over the last few decades, these methods have become increasingly relevant in the study of Quantum Information (QI) theory \cite{nielsen_quantum_2010,bengtsson_zyczkowski_2006}.  These endeavours encompass a spectrum of analyses, ranging from the application of the variational principle to compute the time-optimal evolution and optimal Hamiltonians for quantum systems with specified initial and final states, exemplified by the quantum brachistochrone \cite{carlini2005quantum}, to the determination of optimal quantum circuits by navigating the shortest path in specific curved geometries \cite{nielsen_quantum_2006,nielsen2005geometric, Nielsen_2006}. This trajectory extends further into condensed matter physics, where the application of information-theoretic differential geometry to quantum phase transitions has become a focal point \cite{zanardi_information-theoretic_2007}. Notably, recent scholarly endeavours have extracted geometric insights into quantum-error correcting (QEC) codes, crucial for the development of fault-tolerant quantum computers, stemming from the examination of holographic systems \cite{almheiri_bulk_2015,harlow_ryutakayanagi_2017,pastawski_holographic_2015}.
The net difference between the last set-up and the others is the gap between QFT and quantum mechanics (QM).

This paper has two goals: (i) to sketch a summary of the promising geometrical intuitions given by the works mentioned above, exemplifying the main features; and (ii), to lay down a geometric theory of resources by writing down an {\em action of resource}. Taking as paramount example entanglement as a physical resource, we use it as potential for an action on the manifold of quantum states endowed with the Fubini-Study metric \cite{bengtsson_zyczkowski_2006}. In this way, we can compute the accumulated resource on the path of least action. 
\newline
The paper is structured as follows. In Sec.\ref{ch:AdS/CFT} we describe the insights given by the AdS/CFT correspondence, for a geometrical reinterpretation of the quantum mechanical properties of a theory using the notion of entanglement entropy and the application to quantum computation.
In Sec.\ref{ch:QGT} we pass to highlight the main result of quantum phase transition expressed in terms of the curvature two form, i.e. quantum geometric tensor, for a smooth family of quantum Hamiltonians denoted by a set of parameters.
In Sec.\ref{ch:QCG} quantum computation questions about the smallest numbers of quantum gates required to synthesize a given unitary U are reformulated as the geometrical problem of finding the shortest distance between the identity and U. Such a distance can be computed using the general notion of geodesic upon having defined a proper metric.
In Sec.\ref{ch:Example} we set up a theory in which entanglement is used as a potential for the quantum unitary evolution and we set out to lay down the Euler-Lagrange equations that minimize the distance between the Fubini-Study metric and the entanglement in a given bipartition.
A generalization to other resources is also provided i.e. anti-flatness and coherence. 
We make the explicit example of 2 qubits.
We end with conclusions in Sec.\ref{ch:Conclusions and outlook} where possible outlooks and open questions are made.



\subsection{AdS/CFT: quantum geometrical equivalences}\label{ch:AdS/CFT}
One of the first (unplanned) attempts to give geometric interpretations to some quantum mechanical quantities, over which we have a significant amount of technical control, is undoubtedly the Anti-de Sitter/ Conformal Field Theory (AdS/CFT) conjecture \cite{maldacena1999large,witten_anti_1998}.
The main idea relies on the \textit{correspondence} between theories of quantum gravity (QG) in the $d+1$ dimensions bulk and non-gravitational QFT on the $d$-dimensional boundary, which was derived in the string theory context but it goes far beyond it. 
In its most basic but also precise examples, the correspondence is between theories of QG on AdS background and CFTs. 
There exist 2 versions of this dictionary \cite{witten_anti_1998,gubser1998gauge,banks1998ads,susskind1998holographic} which are tested to be equivalent in the majority of cases \cite{harlow2011operator}, i.e. self-interacting scalar field theory in a fixed AdS background.
Both of them are useful in different contexts and we cite them.

\begin{definition}[GKPW dictionary]
Given the partition function $\mathcal{Z}_{bulk}$ of string theory in $d+1$ dimensions, a function of the boundary conditions of the dynamical fields $\phi=\phi_0$, and the generating functional of correlators in the $CFT_d$ $\mathcal{Z}_{CFT}$, functional of the sources $\phi_0$.
The holographic correspondence reads
\begin{equation}
    \mathcal{Z}_{bulk}[\phi=\phi_0] = \mathcal{Z}_{CFT}[\phi_0] \ ,
\end{equation}
\end{definition}
In the classical limit the correspondence states
\begin{definition}[GKPW classical dictionary]
The classical gravitational action is the generating functional of connected correlators in the CFT
\begin{equation}
    e^{-I_{class}[\phi_0]}= \langle e^{\int \phi_0 \mathcal{O} } \rangle_{CFT}  \ .
\end{equation}
\end{definition}

\begin{definition}[BDHM dictionary]
Bulk operators pulled to the boundary are equivalent to the field theory correlators of the operators dual to the bulk fields.
In Poincarè coordinates it reads
\begin{equation}
    \langle \mathcal{O}(x_1) \ldots   \mathcal{O}(x_n) \rangle_{CFT} = \lim_{z\to 0} z^{-n \Delta}  \langle \phi(x_1,z) \ldots   \phi(x_n,z) \rangle_{bulk} \ .
\end{equation}
\end{definition}
In the context of the above-mentioned AdS/CFT dictionary, a way to address one of the two levels of quantumness, i.e. the entanglement, is the statement made by Ryu and Takayanagi \cite{ryu_holographic_2006}.
The formula proposed asserts that the entanglement entropy of a boundary CFT is equivalent to the area (in Planck units) of a certain minimal surface in the corresponding higher-dimensional gravitational theory within AdS. 
The implications of this formula extend far beyond its original gravitational context and have found deep resonance in QI theory.
The statement goes as follows \cite{harlow_black_2023}
\begin{definition}[Ryu-Takayanagi (RT) formula]\label{def:RT_formula}
Given a CFT state $\rho$ in $d$ dimensions (approximately) dual to a classical bulk geometry with time-reversal symmetry, and a time-reversal invariant bulk Cauchy slice $\Sigma$ with $R\subset \partial \Sigma$ being a spatial region of the boundary CFT, the von Neumann entropy in Planck units on $R$ is given by 
\begin{equation}
S(\rho_R)=\min_\gamma \frac{Area(\gamma)}{4G}~,
\end{equation}
where $\gamma$ varies over all codimension-two surfaces, i.e. ($d-1$)-dimensional surfaces, in $\Sigma$ which are homologous to $R$, meaning there exists a $d$-dimensional homology hypersurface $H \subset \Sigma$ such that $\partial H = R \cup \gamma$. 
Here $G$ denotes the Newton's constant.
The minimal surface $\gamma_R$ is often called RT surface, which is generically not unique.
\end{definition}
Using this definition \ref{def:RT_formula} one can actually recover the Bekenstein-Hawking \cite{bekenstein_black_1972} formula as a peculiar case, where the von Neumann entropy is precisely the thermal entropy and $\gamma_R$ becomes the horizon.
Later works on the subject removed the arbitrary constraint on the time-reversal symmetry, which neglects quantum effects, culminating in the final version of the \textit{quantum extremal surface} (QES) formula \cite{Engelhardt_2015}
\begin{equation}
    S(\rho_R)= \min_\gamma \Big( \ext{\gamma} \Big( \frac{Area(\gamma)+ \ldots}{4G} + S_{bulk}(\rho_H) \Big ) \Big ) ~.
\end{equation}
The augmented entropy is established for a state denoted by $\rho$ within a $CFT_d$ characterized by a semiclassical bulk representation. In simpler terms, there should be a semiclassical expansion in $G$ around a particular state featuring a well-defined geometry, or conceivably, a superposition involving a finite set of such geometries.
Studies on such area laws have been discussed not just in the context of entanglement entropy, but also for Rényi entropies \cite{renyi1961measures} which allow a better understanding of the quantum state of a system due to their fine-grained information and distinguishability power \cite{dong_gravity_2016,dong_holographic_2019}.
It is possible to realize a similar area law in holographic theories for such types of entropies 
\begin{equation}\label{eq: Renyi_RTformula}
    n^2 \partial_n \Big (\frac{n -1}{n} S_n \Big) = \frac{Area(\text{Cosmic Brane}_n)}{4G} ~,
\end{equation}
where $S_n \equiv \frac{1}{1-n} \ln Tr \rho^n$ is a one-parameter generalization of the Von Neumann entropy, recovered by taking $n \to 1$. Here the cosmic brane with subscript $n$ is to be referred to the definition given in \cite{dong_gravity_2016}.
The main result is obtained using the replica trick \cite{lewkowycz2013generalized} in Euclidean signature\footnote{Eulidean continuation may not be always possible, but for the sake of simplicity one can restrict to those cases where this is possible.} where we have direct geometric quantities associated with the close variant of the Rényi entropy in \eqref{eq: Renyi_RTformula}
\begin{equation}\label{eq:GHY_Renyi}
    \partial_n \Big (\frac{n -1}{n} S_n \Big) = \frac{1}{16 \pi G } \int \dd^d x \sqrt{h} \hat{n}^\mu ( \nabla ^\nu \partial_n g_{\mu \nu} - g^{\nu \rho } \nabla_\mu \partial_n g_{\nu \rho}) ~.
\end{equation}
Here the $x^\alpha$ and $h_{\alpha \beta}$ denote the coordinates and the induced metric respectively.
It is important to notice that the right-hand side of \eqref{eq:GHY_Renyi} is essentially the derivative concerning $n$ of the Gibbons-Hawking-York (GHY) term obtained by the Einstein-Hilbert action, which in general accounts for the well-posedness of the theory.
In this picture, Rényi entropies trace a path to a fine-grained version and interpretation of quantum mechanical quantities from their geometrical duals. 
\begin{figure}
    \centering
    \includegraphics[width=0.7\columnwidth]{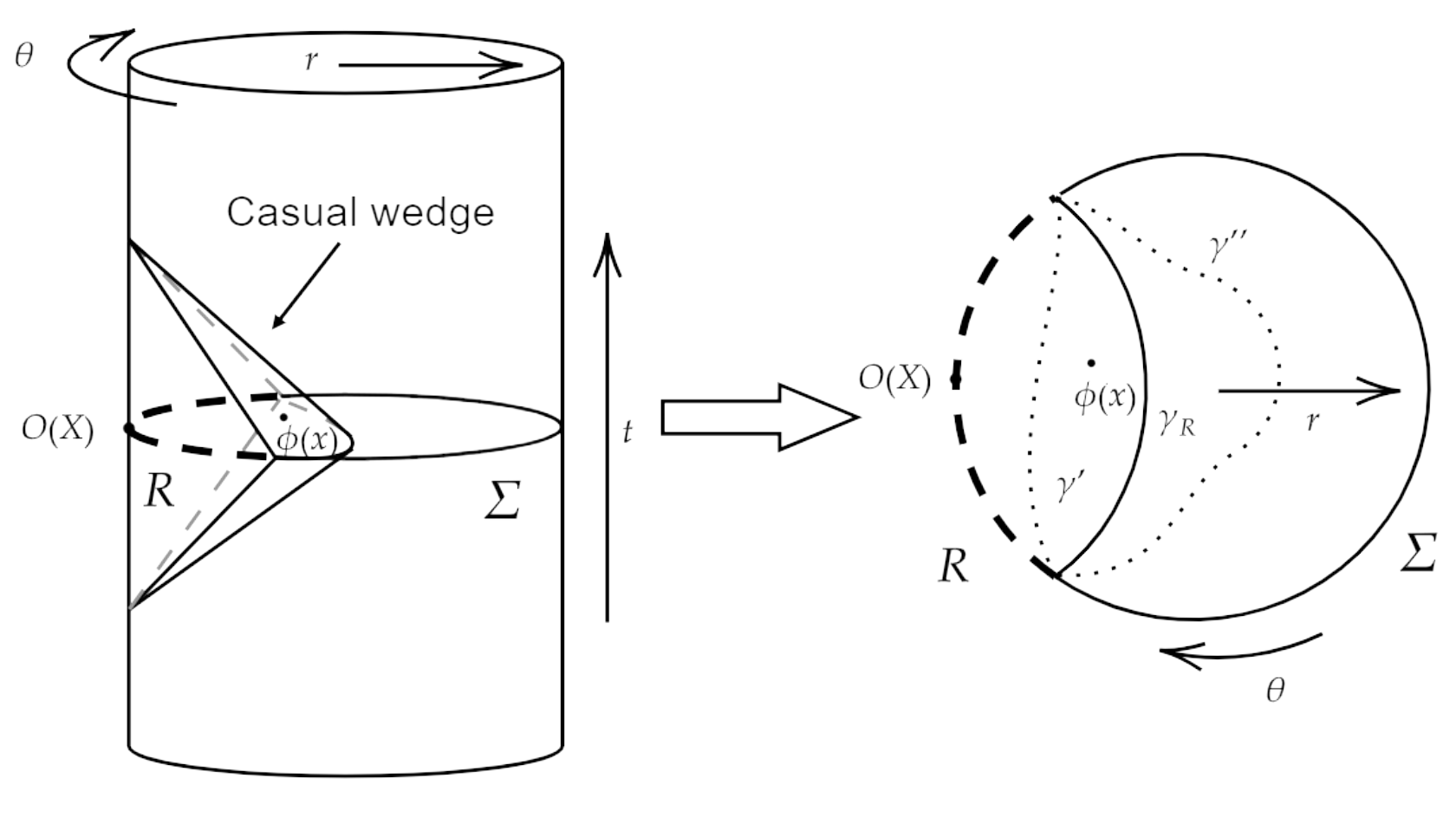}
    \caption{Example of bulk field reconstruction in the casual wedge. We highlighted the general structure of the area defined by the RT surface. }
    \label{fig:AdS/CFT_RT entropy}
\end{figure}

More recently, the AdS/CFT correspondence offers a way to re-think QI questions such as quantum error correction (QEC) codes \cite{almheiri_bulk_2015,harlow_ryutakayanagi_2017,pastawski_holographic_2015}, essential for the quantum advantage. 
The major feature lies in the AdS-Rindler reconstruction \cite{banks1998ads,Hamilton:2006az}, which plays a pivotal role in understanding the holographic principle. 
This technique allows mapping quantum states and operators in the CFT, particularly those within the Rindler wedge, to their corresponding counterparts in the bulk AdS space, see Fig. \ref{fig:AdS/CFT_RT entropy}.
It unveils the holographic nature of the AdS/CFT duality, where the bulk theory is encoded on the boundary.
The main idea behind the reformulation of QEC in the context of AdS/CFT, using the BDHM dictionary, is that local bulk operators can be viewed as logical operations that act upon an encoded subspace. 
As these operators are shifted deeper into the radial direction, they progressively enhance the protection of the subspace against errors localized near the boundary. 
This framework uses holographic codes to protect QI from errors and decoherence in the bulk AdS space.


\subsection{Quantum geometric tensors}\label{ch:QGT}
In general relativity (GR), it is important to identify whether a point in the manifold is regular despite the metric blowing up at infinity\footnote{This holds in the more general framework of differential geometry as well.}.
Many of these points are entirely regular, and the infinities arise solely due to using an unsuitable coordinate system. 
Hence, one seeks the regularity of the scalar quantities, each for every independent component of the Riemann tensor, e.g. Ricci scalar, Kretschmann scalar, etc.
These kinds of quantities encapsulate much information about the theory at hand. 
Similarly is possible to exploit the tools of Riemann geometry for quantum mechanics (QM) in studies such as quantum phase transitions.

\begin{figure}
    \centering
    \includegraphics[width=0.5\columnwidth]{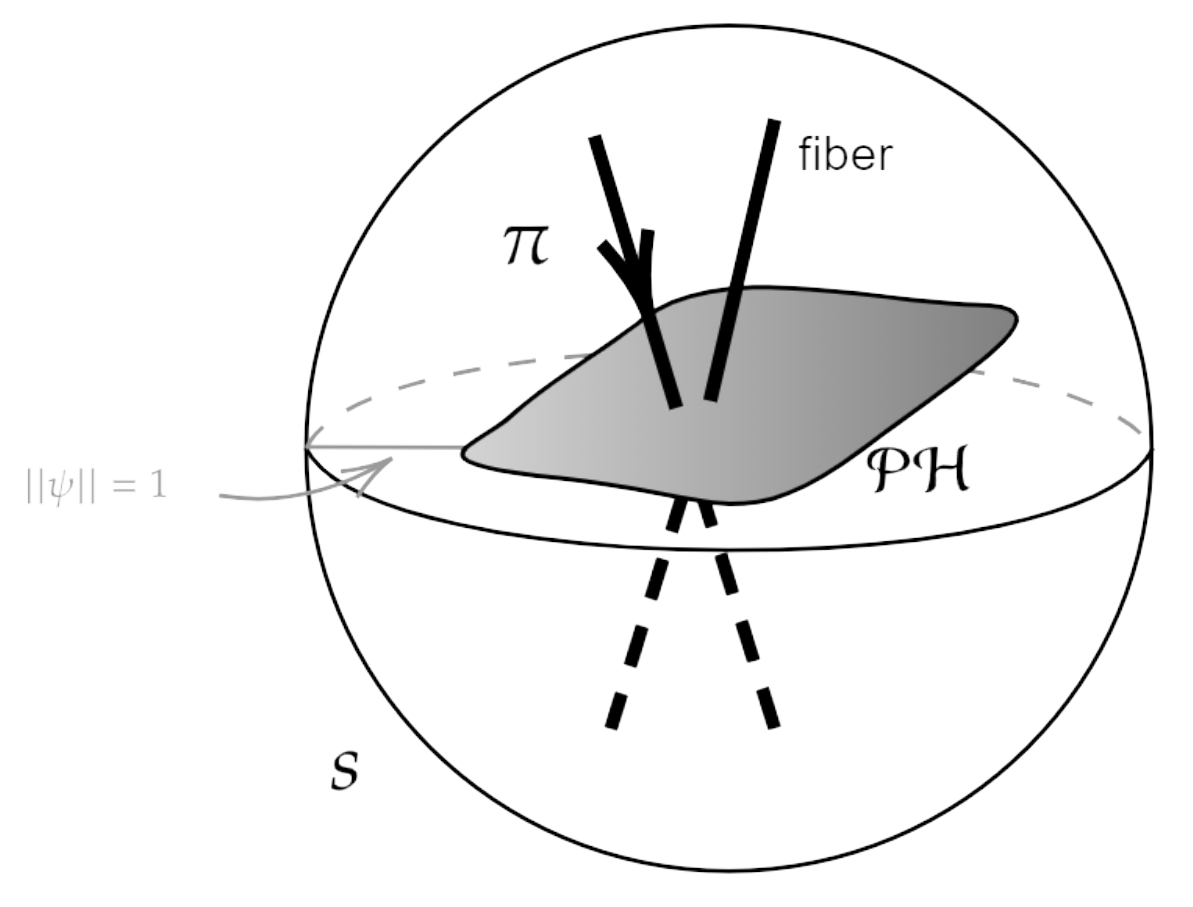}
    \caption{A pictorial illustration of a principal fiber bundle with total space the unit ball $S:= \{ \ket{\psi} \in \mathcal{H} | \norm{\psi}=1 \}$ of the Hilbert space $\mathcal{H}$ with base space $\mathcal{PH}$ and projective map $\pi: S \to \mathcal{PH} / \ket{\psi} \to \{ e^{i \theta} \ket{\psi} | \theta \in [0,2\pi) \} $.}
    \label{fig:principal fiber bundle}
\end{figure}

Given a smooth family of quantum Hamiltonians $H(\lambda)$, characterised by a set of parameters $\lambda \in \mathbb{R}^m$ which form a manifold $\mathcal{M}$, we can find a ground state, considered unique for simplicity, namely one can define a map $\Psi_0$ that goes from the parameter manifold to the associated ground state.
Among the required physical properties imposed by QM there is the fact one cannot observe the phase of a vector state\footnote{Although Berry \cite{berry1984quantal} noted that an adiabatic change in the system's phase could have observable consequences, such as the Aharonov-Bohm effect where the magnetic field enclosed by two interference paths is the adiabatic parameter.}, hence one aims to study the projective space of the Hilbert-space $\mathcal{PH}$ where the vector states are normalized, see Fig. \ref{fig:principal fiber bundle}.
This space comes with a metric \cite{provost1980riemannian} (see \ref{app:QGT}), i.e. the Fubini-Study distance, which correctly quantifies the statistical distinguishability between two pure quantum states.
So, under the pull-back of  $\Psi_0$ one can recover the metric in the parameter manifold $\mathcal{M}$.
Such a metric reads \cite{zanardi_information-theoretic_2007}
\begin{equation}
    g_{\mu \nu}= \Re \sum_{n \neq 0} \frac{\bra{\Psi_0(\lambda)} \partial_\mu H \ket{\Psi_n(\lambda)} \bra{\Psi_n(\lambda)} \partial_\nu H \ket{\Psi_0(\lambda)}}{(E_n(\lambda)-E_0(\lambda))^2}~,
\end{equation}
where the first order perturbative expansion has been used on the ground state $\ket{\Psi_0(\lambda + \delta \lambda)}$, $E_n(\lambda)$ is the energy of the $n$-th state linked to the parameter manifold and $\partial_\mu := \frac{\partial}{\partial \lambda^\mu}$ with $\mu= 1,\ldots, dim \mathcal{M}$.

Upon studying the scalar curvature of the metric after proper re-scaling, we can observe the presence or not of divergences in the parameter manifold, which indicates the existence of a phase transition.
Thus, the distance between ground states can be used to measure proximity to a quantum phase transition. 
At the critical point, the distance is small, indicating that the ground states are similar. 
As the control parameters are moved away from the critical point, the distance increases, indicating that the ground states differ significantly.
Also away from equilibrium, it is possible to recover important geometric information highlighting the powerful tool of geometric tensors \cite{rattacaso2020quantum}. 

In conclusion, in order to have a good self-contained introduction to the topic we suggest the reader consult \cite{zanardi_information-theoretic_2007} and references therein.


\subsection{Quantum computation and geometric approach}\label{ch:QCG}
In the realm of quantum computation, a key challenge is to discover effective quantum circuits capable of creating specific unitary operations.
Almost a decade ago it was proposed to introduce a geometrical approach to finding guiding principles for such a challenge \cite{nielsen_quantum_2006, nielsen2005geometric, Nielsen_2006, Nielsen2006, dowling2008geometry}.

Similar to what was described in the section \ref{ch:QGT}, a Riemannian metric is to be determined on the space of $n$-qubit unitary operations.
This metric has to quantify the number of quantum gates required to synthesize a desired unitary $U$ by measuring the distance from the identity operation $I$.
In this way, one understands the intrinsic difficulty in the synthesis of $U$ as well.
The idea is to consider the group $SU(2^n)$ of $n$-qubit unitary operations as manifolds so that at each point a tangent vector, i.e. an element of the associated Lie algebra, can be defined and interpreted as a traceless Hamiltonian $H$.
The same argument can be applied to $U(2^n)$.
Hence, the Riemannian metric $\langle\cdot,\cdot\rangle_U$ at a point $U$ is a positive-definite bilinear form and for the sake of simplicity one can consider it as a right-invariant metric, i.e. it is constant as a function of $U$.
A metric of this type, i.e. the standard metric, is described in \cite{dowling2008geometry} for $H,J \in su(2^n)$
\begin{equation}\label{eq:metric_computing}
    \langle H ,J \rangle :=  \frac{\Tr(H \mathcal{P}(J))+q \Tr(H \mathcal{Q}(J)) }{2^n} ~,
\end{equation}
where $su(2^n)= \mathcal{P} + \mathcal{Q}$ and $\mathcal{P}$ represent the subspace of $n$-qubit Hamiltonians which contains only 1- and 2-body terms, whereas $\mathcal{Q}$ is its complementary. 
Here $q$ is a penalty parameter necessary for later considerations.

From the standard metric \eqref{eq:metric_computing} an induced distance $d(\cdot,\cdot)$ can be found between two points $U,V \in SU(2^n)$ on the manifold. 
Given a curve $U(t)$ generated by the Hamiltonian $H(t)$ according to $\dot{U}=- i H U$, $d(\cdot,\cdot)$ is defined by the minimal length, i.e. $\int \dd t \sqrt{\langle H(t) ,H(t) \rangle}  $, of any curve joining the initial $U(t_0)=U$ and the final $U(t_f)=V$ point.
Such a distance between an arbitrary unity $U$ and the identity $I$, $d(I,U)$, is the desired quantity mentioned above.

\begin{figure}
    \centering
    \includegraphics[width=0.6\columnwidth]{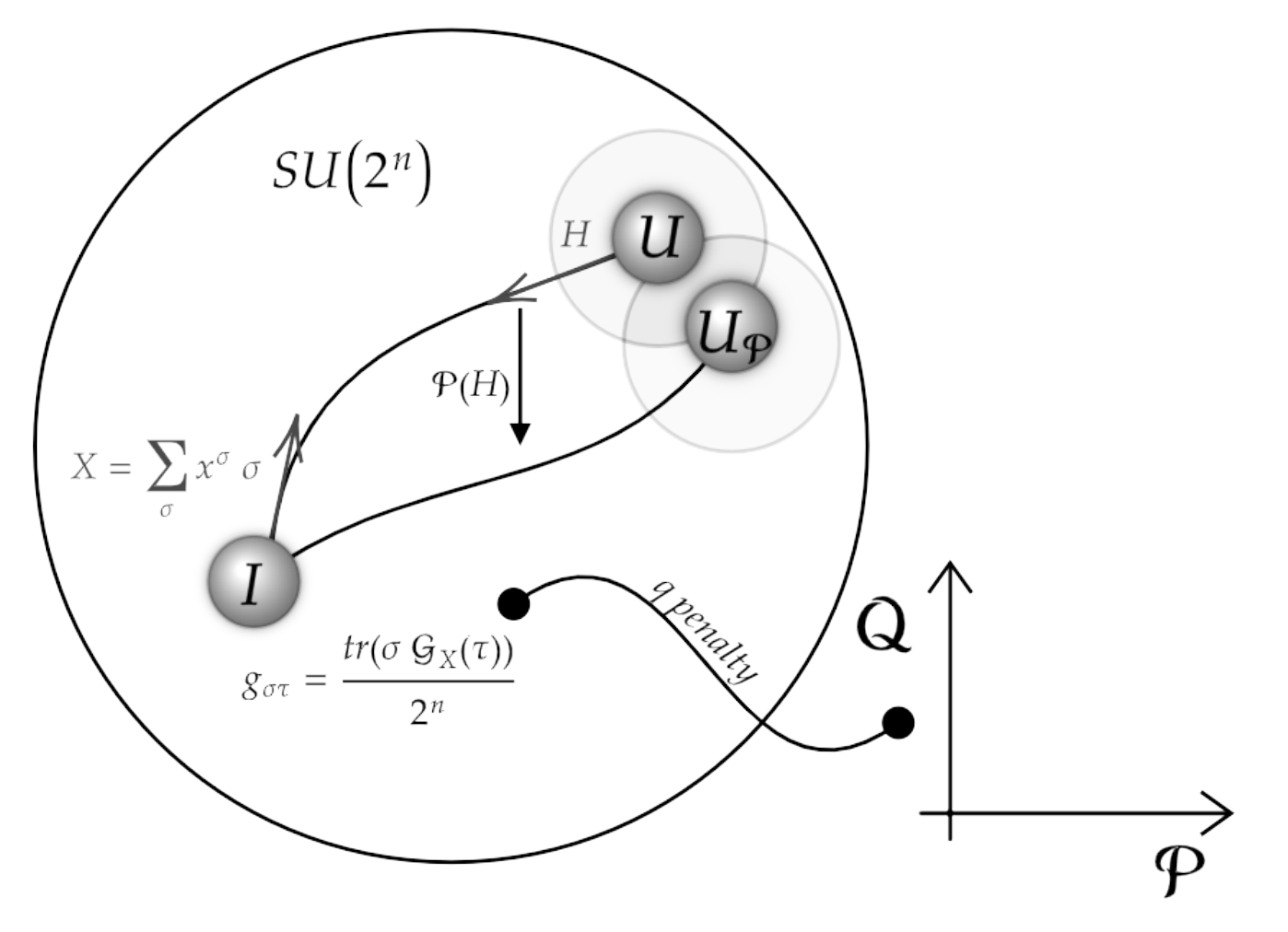}
    \caption{The pictorial representation of the distance between the unitary $U$ and the identity $I$ for $SU(2^n)$ with the penalty metric $g_{\sigma \tau}$ and the tangent vector expressed in Pauli local coordinates.}
    \label{fig:Quantum computing}
\end{figure}


\section{Accumulation of a resource along a minimal path}\label{ch:Example}
Now, we use some of the machinery of the spectrum of geometrical methods we presented to study the accumulation of entanglement along a given path in the manifold of parameters that describe the system.
Consider a Hamiltonian $H$, which depends, as in Sec.\ref{ch:QGT}, by a set of parameters $\lambda \in \mathbb{R}^m$.
The change of these parameters in the manifold they span can be viewed as a curve $\lambda \equiv \lambda(s)$ with $s \in [0,1]$.
A simple question can be asked: What is the least expensive path, in terms of a given resource, with fixed endpoints $\lambda_A$ and $\lambda_B$ for a give unitary evolution of a reference state $\ket{\Omega}$ of the type:
\begin{equation}
    \ket{\psi(\lambda(s))}= U(\lambda(s)) \ket{\Omega} ~,
\end{equation}
with $U(\lambda(s))= e^{i H(\lambda(s))}$?
Here we define the density operator associated with such a state as
\begin{equation}
    \psi(s) = \ket{\psi(s)} \bra{\psi(s)}= U(s) \Omega U^\dag(s)~,
\end{equation}
where we used a shortened notation for the dependence on $s$ and $\Omega=\ket{\Omega}\bra{\Omega}$.
Notice that $U$ is not a time evolution of the Hamiltonian, but just the standard way to rewrite a unitary evolution given some hermitian operator such as $H$.

Now, minimising a path for a resource can be stated as the standard variational principle for the minimization of an action
\begin{equation}\label{eq:Action_classical}
    I[\lambda(s), \lambda'(s)]=\int_0^1 ds ~ L(\lambda(s), \lambda'(s))~.
\end{equation}
Here $\lambda'(s)=\frac{d \lambda}{ds}$ and $L$ is the Lagrangian of the system.
The Lagrangian may be constructed as the usual difference between kinetic and potential terms. 
The former is the line element of the Fubini-Study distance (Sec.\ref{ch:QGT}), whereas the latter is the entanglement in a given bipartition of the Hilbert space
\begin{equation}\label{eq:Lagrangian}
    L(\lambda(s),\lambda'(s))= K_1-V :=\sqrt{\bra{\psi'}(\mathds{1}-\psi)\ket{\psi'}}- (1-\tr_X[\psi_X^2]) ~,
\end{equation}
with $\ket{\psi'(s)}=\frac{d}{ds}\ket{\psi(s)}= U'(s) \ket{\Omega}$ and $\psi_X(s)= \tr_Y[\psi(s)]$ being the partial trace over a subsystem of the whole Hilbert space assuming the factorization $\mathcal{H}=\mathcal{H}_X \otimes \mathcal{H}_Y$.
Computing the Euler-Lagrange (EL) equations for $\lambda^\mu$
\begin{equation}
    \frac{d}{ds} \frac{\partial L}{\partial \lambda'^\mu}=\frac{\partial L}{\partial \lambda^\mu}  ~,
\end{equation}
they can be expressed in the following equivalent form:
\begin{align}\label{eq:EL_FS}
    &2 \tr[(\mathds{1}-\psi)\Tilde{\psi}] \tr[\psi'(\partial_{\lambda'} \Tilde{\psi} )-(\partial_\lambda \psi) \Tilde{\psi}+(\mathds{1}-\psi) \Big(\partial_\lambda \Tilde{\psi}-\frac{d}{ds}(\partial_{\lambda'} \Tilde{\psi} ) \Big)]\nonumber \\
    &+4 \tr^{3/2}[(\mathds{1}-\psi)\Tilde{\psi}] \tr_X[\partial_\lambda(\psi_X ) \psi_X +\psi_X \partial_\lambda \psi_X] + \tr [(\mathds{1}-\psi) \partial_{\lambda'} \Tilde{\psi} ]\tr[(\mathds{1}-\psi)\Tilde{\psi}' - \psi' \Tilde{\psi} ] =0~.
\end{align}
A compact notation has been used, namely $\Tilde{\psi}=U' \Omega U'^\dag$, $\partial_{\lambda'}= \frac{\partial}{\partial \lambda'^\mu}$, $\partial_{\lambda}= \frac{\partial}{\partial \lambda^\mu}$ and as before $':= \frac{d}{ds}$.
The solution $\lambda(s)$ of the EL equations above stabilizes the action, $\delta I=0$. One can then proceed to compute the accumulated resource along this path, namely
\begin{align}\label{eq:entanglement_acc}
    \overline{E} = \int_0^1 ds ~ E(\psi(s))
\end{align}
with $E(\psi(s))=1-\tr_X[\psi_X^2]$.
The first term in \eqref{eq:Lagrangian} given by the Fubini-Study alone would find the geodesics on the manifold of quantum states. In our setting, the resource $E$ represents a potential, so we do not have a `free falling' particle, but the potential determines a geodesic deviation. However, the kinetic term can also be substituted by its square, as they define the same geodesics \cite{nielsen2005geometric}. This will allow us to obtain much simpler differential equations. Indeed, by defining the Lagrangian
\begin{equation}\label{eq:Lagrangian2}
    L(\lambda(s),\lambda'(s))= K_2-V :={\bra{\psi'}(\mathds{1}-\psi)\ket{\psi'}}- (1-\tr_X[\psi_X^2]) ~,
\end{equation}
we obtain the much simpler EL equations
\begin{equation}\label{eq:EL_FS2}
    \tr[(\mathds{1}-\psi)\partial_\lambda \Tilde{\psi} - (\partial_\lambda \psi)\Tilde{\psi}]+ \tr_X[\partial_\lambda(\psi_X ) \psi_X +\psi_X \partial_\lambda \psi_X]+ \tr[\psi' \partial_{\lambda'}\Tilde{\psi}-(\mathds{1}-\psi)\frac{d}{ds}(\partial_{\lambda'} \Tilde{\psi})]=0~,
\end{equation}
with same conventions as before.

Since we defined the Lagrangian in \eqref{eq:Action_classical} as the competition between a kinematic and a potential term, nothing prohibits choosing a different potential such as 
\begin{equation}\label{eq:flatness}
    F(\psi)=\tr[\psi_X^3]-\tr^2[\psi_X^2]~,
\end{equation}
that quantifies the \textit{anti-flatness} of the spectrum of $\psi_X$ \cite{tirrito2023quantifying}.
The EL equations are very similar to those obtained before for the entanglement \eqref{eq:EL_FS} and \eqref{eq:EL_FS2} except for the potential part.
For the Fubini-Study kinetic term $K_1$ one has
\begin{align}\label{eq:EL_FS3}
    &2 \tr[(\mathds{1}-\psi)\Tilde{\psi}] \tr[\psi'(\partial_{\lambda'} \Tilde{\psi} )-(\partial_\lambda \psi) \Tilde{\psi}+(\mathds{1}-\psi) \Big(\partial_\lambda \Tilde{\psi}-\frac{d}{ds}(\partial_{\lambda'} \Tilde{\psi} ) \Big)]\nonumber \\
    &+4 \tr^{3/2}[(\mathds{1}-\psi)\Tilde{\psi}] (-  \tr_X[\sum_{k=1}^3 \psi_X^{k-1} \partial_\lambda (\psi_X) \psi_X^{3-k}]+ 2 \tr_X[\psi_X^2] \tr_X[\partial_\lambda(\psi_X ) \psi_X +\psi_X \partial_\lambda \psi_X]) \nonumber\\
    &+ \tr [(\mathds{1}-\psi) \partial_{\lambda'} \Tilde{\psi} ]\tr[(\mathds{1}-\psi)\Tilde{\psi}' - \psi' \Tilde{\psi} ] =0~,
\end{align}
whereas for its square $K_2$
\begin{align}\label{eq:EL_FS4}
   &\tr[(\mathds{1}-\psi)\partial_\lambda \Tilde{\psi} - (\partial_\lambda \psi)\Tilde{\psi}]+ (-  \tr_X[\sum_{k=1}^3 \psi_X^{k-1} \partial_\lambda (\psi_X) \psi_X^{3-k}]+ \nonumber \\
   &2 \tr_X[\psi_X^2] \tr_X[\partial_\lambda(\psi_X ) \psi_X +\psi_X \partial_\lambda \psi_X])+ \tr[\psi' \partial_{\lambda'}\Tilde{\psi}-(\mathds{1}-\psi)\frac{d}{ds}(\partial_{\lambda'} \Tilde{\psi})]=0~.
\end{align}
Even in this case, it is possible to evaluate the accumulated resource along the minimal path
\begin{align}\label{eq:flatness_acc}
    \overline{F} = \int_0^1 ds ~ F(\psi(s))~.
\end{align}
It is with minimum effort that one can also think to compute the accumulated entanglement along the minimal path with respect to the Lagrangian containing the anti-flatness and vice versa.

Another important resource in quantum physics is {\em coherence}. All quantum properties are closely related to quantum coherence, including uncertainty, contextuality, and entanglement. Experimentally coherence manifests in interference and quantum fluctuations. It is at the roots of quantum advantage in quantum information processing protocols \cite{streltsov2017colloquium}.

The (squared) 2--norm of coherence \cite{de2016genuine,zanardi2017coherence,korzekwa2018coherifying} is defined as
\begin{equation}
    c_{2,B}(\sigma) = \left\| (\mathcal I -\mathcal D_B) \sigma  \right\|_2^2  = \mbox{Pur}(\sigma) -\mbox{Pur} \left[ \mathcal D_B (\sigma) \right]
\end{equation}
where $\left\| X \right\|_2 \coloneqq \sqrt{\Tr \left( X^\dagger X \right)}$ denotes the (Schatten) 2-norm, $\mathcal D_{B_S} (X) \coloneqq \sum_k \chi_k X \chi_k$ is the dephasing superoperator in the basis $B_S =\{ \chi_k \}_{k=1}^{d_S}$  of \text{rank-1} orthogonal projectors $\chi_k=\ket{k}\!\bra{k}$, and finally
$\mbox{Pur}(\rho) \coloneqq \Tr (\rho^2)$ denotes the purity functional. We can thus define the new potential for the pure states $\psi(s)$
\begin{equation}\label{eq:coherence}
    Q(\psi)=1-\mbox{Pur} \left[ \mathcal D_B (\psi) \right]
\end{equation}
The associated EL equations for the two choices of the kinetic term $K_1, K_2$ are
\begin{align}\label{eq:EL_FS_De}
    &2 \tr[(\mathds{1}-\psi)\Tilde{\psi}] \tr[\psi'(\partial_{\lambda'} \Tilde{\psi} )-(\partial_\lambda \psi) \Tilde{\psi}+(\mathds{1}-\psi) \Big(\partial_\lambda \Tilde{\psi}-\frac{d}{ds}(\partial_{\lambda'} \Tilde{\psi} ) \Big)]\nonumber \\
    &+4 \tr^{3/2}[(\mathds{1}-\psi)\Tilde{\psi}] \tr[\psi \mathcal D_B (\partial_\lambda \psi )  + \partial_\lambda \psi \mathcal  D_B (\psi) ) ] + \tr [(\mathds{1}-\psi) \partial_{\lambda'} \Tilde{\psi} ]\tr[(\mathds{1}-\psi)\Tilde{\psi}' - \psi' \Tilde{\psi} ] =0~.
\end{align}
and 
\begin{align}\label{eq:EL_FS_De_2}
   &\tr[(\mathds{1}-\psi)\partial_\lambda \Tilde{\psi} - (\partial_\lambda \psi)\Tilde{\psi}+\psi \mathcal D_B (\partial_\lambda \psi )  + \partial_\lambda \psi \mathcal  D_B (\psi) )  +\psi' \partial_{\lambda'}\Tilde{\psi}-(\mathds{1}-\psi)\frac{d}{ds}(\partial_{\lambda'} \Tilde{\psi})]=0~.
\end{align}

Once again the accumulation is defined as
\begin{equation}
    \label{eq:coherence_acc}
    \overline{Q} = \int_0^1 ds ~ Q(\psi(s))~.
\end{equation}

In the next section, we show how these equations can be integrated (numerically) with much greater ease.

\subsection{2 qubit example}
Let us start with the case of straightforward implementation.
Define the complete basis computational basis $\{\ket{ij}\}_{i,j=0,1}$ and the usual Pauli operators $P=\{\mathds{1},X,Y,Z\}$.
Then the hermitian operator $H$, which defines the unitary evolution operator $U=e^{iH(\theta(s),\phi(s))}$ and the reference state are
\begin{equation}
    \ket{\Omega}=\frac{\ket{00}+\ket{01}}{\sqrt{2}}~,\quad H(\theta,\phi)=\theta X \otimes X + \phi Z \otimes Z~.
\end{equation}
So we can properly compute the state under unitary evolution
\begin{align}
    \ket{\psi(s)}&= e^{iH(s)}\ket{\Omega} \nonumber\\
    &=\frac{1}{\sqrt{2}}(e^{i \phi(s)}\cos \theta(s) \ket{00}+e^{-i \phi(s)}\cos \theta(s) \ket{01}\\
    &+i e^{-i \phi(s)} \sin \theta(s) \ket{10}+ i e^{i \phi(s)} \sin{\theta(s)} \ket{11})~.
\end{align}
To have a well-posed problem we fix the endpoints such as
\begin{equation}
    \theta(0)=\phi(0)=0, \quad \theta(1)=\frac{\pi}{4}, \quad \phi(1)=2\pi ~.
\end{equation}
Notice that for $s=0$ we obtain the separable state $\ket{\Omega}$ and for $s=1$ we have the separable state
\begin{equation}
    \ket{\psi(s=1)}=\frac{1}{2}(\ket{0}+i\ket{1}) ( \ket{0}+ \ket{1})~.
\end{equation}
Therefore the Lagrangian in \eqref{eq:Lagrangian} is
\begin{equation}
    L(\theta(s),\phi(s),\theta'(s),\phi'(s))= \sqrt{\theta '(s)^2+\phi '(s)^2}-\frac{1}{2} \sin ^2(2 \theta (s)) \sin ^2(2 \phi (s)) ~,
\end{equation}
with the following EL equations
\begin{align}\label{eq:EL1}
   \sin (4 \theta (s)) \sin ^2(2 \phi (s))+ \frac{\theta ''(s) \phi '(s)^2-\theta '(s) \phi '(s) \phi ''(s)}{\left(\theta '(s)^2+\phi '(s)^2\right)^{3/2}}&=0 ~,\\
   \label{eq:EL11}
    \sin ^2(2 \theta (s)) \sin (4 \phi (s)) +\frac{\theta '(s)^2 \phi ''(s)-\theta '(s) \theta ''(s) \phi '(s)}{\left(\theta '(s)^2+\phi '(s)^2\right)^{3/2}}&=0 ~.
\end{align}
Hence, the mean entanglement for the solution that stabilizes the action reads
\begin{equation}
    \overline{E}= \int_0^1 ds~ \frac{1}{2} \sin ^2(2 \theta (s)) \sin ^2(2 \phi (s))~, 
\end{equation}
As we mentioned in the previous section, since \eqref{eq:EL1}-\eqref{eq:EL11} do not have straightforward solutions, we pass to exploit the $K_2$ kinetic term \eqref{eq:Lagrangian2}.
The EL equations in this case reads
\begin{align}
\label{eq:EL_lin1}
    \sin (4 \theta (s)) \sin ^2(2 \phi (s))+ 2 \theta ''(s) &=0~,\\
    \label{eq:EL_lin2}
    \sin ^2(2 \theta (s)) \sin (4 \phi (s))+2 \phi ''(s)&=0~.
\end{align}
which can be integrated numerically to find the stabilizing path. Finally, the entanglement \eqref{eq:entanglement_acc}, the anti-flatness \eqref{eq:flatness_acc} and the coherence \eqref{eq:coherence_acc} accumulated along this optimal path can be computed numerically, obtaining
\begin{equation}
    \overline{E}= \int_0^1 ds~ \frac{1}{2} \sin ^2(2 \theta (s)) \sin ^2(2 \phi (s))\simeq 0.131526 ~,
\end{equation}
\begin{align}
    \overline{F}= \int_0^1 ds~ \frac{1}{64} & (-2 \sin ^4(2 \theta (s)) \cos (8 \phi (s))-2 \sin ^2(4 \theta (s)) \cos (4 \phi (s))+ \nonumber \\
    &+\sin ^2(2 \theta (s)) (3 \cos (4 \theta (s))+5)) \simeq 0.0281002 ~,
\end{align}
\begin{equation}
    \overline{Q}= \int_0^1 ds~ \frac{1}{8} (5-\cos (4 \theta (s))) \simeq 0.631313 ~,
\end{equation}

As a remark, notice that if we just want to transform unitarily a separable state into a separable state, we can do it without accumulating any entanglement along the way, by suitable single-qubit rotations. 
This would correspond to a trivial action without any competition between the kinetic term and the potential term. 
In our setting, the action is minimized taking into account that there is a geodesic structure on the manifold of quantum states. 
This results in an accumulated entanglement in the optimal path between two separable states and makes the action theory of resource non-trivial.

We, now repeat the same procedure for the following Lagrangian
\begin{align}
     L(\theta(s),\phi(s),\theta'(s),\phi'(s))&= {\bra{\psi'}(\mathds{1}-\psi)\ket{\psi'}}-(\tr[\psi_X^3]-\tr^2[\psi_X^2])\nonumber \\
     &= \theta '(s)^2+\phi '(s)^2-\frac{1}{64} (-2 \sin ^4(2 \theta (s)) \cos (8 \phi (s))+ \nonumber \\
     &-2 \sin ^2(4 \theta (s)) \cos (4 \phi (s)) +\sin ^2(2 \theta (s)) (3 \cos (4 \theta (s))+5)) ~,
\end{align}
where now the accumulated resources along the extremal path read
\begin{equation}\label{eq:E_F_mean_FS4}
    \overline{E}\simeq 0.125793 ~, \quad \overline{F}\simeq 0.0275045 ~, \quad  \overline{Q}\simeq 0.62578 ~ .
\end{equation}
The above numerical results \eqref{eq:E_F_mean_FS4} are obtained with respect to the following EL equations
\begin{align}
    &2 \theta ''(s)-\frac{1}{4} \sin ^3(2 \theta (s)) \cos (2 \theta (s)) \cos (8 \phi (s))-\frac{1}{8} \sin (8 \theta (s)) \cos (4 \phi (s))+\frac{1}{16} \sin (4 \theta (s))+\frac{3}{32} \sin (8 \theta (s))=0  ~,\\
    &\frac{1}{4} \sin ^4(2 \theta (s)) \sin (8 \phi (s))+\frac{1}{8} \sin ^2(4 \theta (s)) \sin (4 \phi (s))+2 \phi ''(s) =0 ~.
\end{align}

The last Lagrangian we want to consider is the one involving the coherence  Eq.\eqref{eq:coherence} 
\begin{align}
     L(\theta(s),\phi(s),\theta'(s),\phi'(s))&= {\bra{\psi'}(\mathds{1}-\psi)\ket{\psi'}}-(1-\mbox{Pur} \left[ \mathcal D_B (\psi) \right])\nonumber \\
     &= \theta '(s)^2+\frac{1}{8} (\cos (4 \theta (s))-5)+\phi '(s)^2) ~,
\end{align}
which EL equations are
\begin{align}
    2 \theta ''(s)+\frac{1}{2} \sin (4 \theta (s)) &=0~,\\
    2 \phi ''(s) &=0~.
\end{align}
Hence the accumulation of the above resources is
\begin{equation}
    \overline{E} \simeq0.131308 ~, \quad \overline{F}\simeq 0.0280876 ~, \quad \overline{Q}\simeq 0.631308~.
\end{equation}


\section{Conclusions}\label{ch:Conclusions and outlook}

The theory of action of resources sketched in this paper is far from being complete. It is a first tentative to connect quantum resources (e.g., entanglement, anti-flatness or coherence), to the geometry of quantum states. In this way, one is going beyond the idea that quantum evolution is generated by a Hamiltonian with a fixed graph of interactions, which are dictated by a given space (and time) in which the graph is embedded. Possibly, the path that minimizes the action of entanglement could be reinterpreted as the one generated by a local Hamiltonian, thereby making the graph structure emerge. In perspective, there is a large body of work to be done in this direction which will be leading our future endeavors. For example, it would be interesting to study the gap of accumulated resource between the optimal path of the action of resource and the time evolution induced by other actions, e.g., the brachistochrone with appropriate constraints \cite{carlini2005quantum,rezakhani_quantum_2009}. Another interesting direction is to see if, starting from a complete graph for the interaction Hamiltonian, the optimal path of resource does select which degrees of freedom are most entangled with which, and use this as a definition of an emergent graph distance. This would require a thorough study of multi-party entanglement \cite{facchi2010multipartite}. We hope that the community may find these possibilities equally interesting.

\section{Aknowledgements}
A.H. acknowledges financial support from PNRR MUR project PE0000023-NQSTI and PNRR MUR project CN 00000013-ICSC.


\appendix
\section{Meaningful metric tensor}\label{app:QGT}
Here we explicitly compute the derivation of the Riemannian structure induced by the hermitian product on the projective Hilbert space $\mathcal{PH}$ with a meaningful metric tensor on any manifold of quantum states \cite{provost1980riemannian}.

Let us start with a family of normalized vectors $\{ \psi (\lambda) \}_{\norm{\psi}=1}$ of some Hilbert space $\mathcal{H}$ that depends on a $m$-dimensional parameter $\lambda \in \mathbb{R}^m$.
The metric induced by the distance between two vectors in the family is recovered as follows, up to second order in Einstein summation convation:
\begin{align}
    \norm{\delta \psi}^2 :=\norm{\psi(\lambda + \dd \lambda)-\psi(\lambda)}^2 & = \braket{\psi(\lambda + \dd \lambda)-\psi(\lambda)}{\psi(\lambda + \dd \lambda)-\psi(\lambda)}\\
    &= \braket{\partial^i \psi}{ \partial^j \psi } \dd \lambda_i \dd \lambda_j ~, \label{eq: module}
\end{align}
where $\partial_\mu := \frac{\partial}{\partial \lambda^\mu}$ with $\mu= 1,\ldots, dim \mathcal{M}$ and we suppressed the $\lambda$ notation in the second line.
Here we will use the fact that the $\braket{\cdot}{\cdot}$ is a sesqui-linear inner product defined on $\mathcal{H}$, having in this sense a complete metric space.
We used the Taylor expansion
\begin{equation}
    \psi(\lambda + \dd \lambda)= \psi(\lambda ) + \partial^i \psi(\lambda ) \dd \lambda_i  + \frac{1}{2} \partial^{ij} \psi(\lambda ) \dd \lambda_i \dd \lambda_j ~.
\end{equation}
By defining the \eqref{eq: module} as
\begin{equation}
    \braket{\partial_i \psi}{ \partial_j \psi }= \gamma_{ij} + i \sigma_{ij} ~,
\end{equation}
consisting of a symmetric real part and an antisymmetric imaginary part, we have that
\begin{align}
     \norm{\delta \psi}^2&= \gamma^{ij} \dd \lambda_i \dd \lambda_j + \frac{1}{2} i (\sigma^{ij} \dd \lambda_i \dd \lambda_j + \sigma^{ji} \dd \lambda_j \dd \lambda_i ) \\
     &= \gamma^{ij} \dd\lambda_i \dd \lambda_j + \frac{1}{2} i (\sigma^{ij} \dd \lambda_i \dd \lambda_j - \sigma^{ij} \dd \lambda_i \dd \lambda_j ) \\
     &= \gamma^{ij} \dd\lambda_i \dd \lambda_j  ~.
\end{align}
It is straightforward to determine what happens when we have a gauge invariance related to the total phase of a vector state as discussed in \ref{ch:QGT}.
The manifold of rays does not distinguish two points up to a total phase leading to
\begin{equation}
    \psi'(\lambda)=e^{i \alpha(\lambda)} \psi (\lambda)~, \quad \partial_i \psi'(\lambda)=e^{i \alpha(\lambda)} [i (\partial_i \alpha) \psi (\lambda) + \partial_i \psi (\lambda) ]~.
\end{equation}
Hence, we would like that the metric tensors for $\{ \psi ' (\lambda) \}_{\norm{\psi'}=1}$ and $\{ \psi (\lambda) \}_{\norm{\psi}=1}$ with a smooth $\alpha(\lambda)$ are equal.
Namely, we want to address the fact that
\begin{align}\label{eq:transormation_metric}
    \gamma'_{ij}& = \Re [\braket{\partial^i \psi'}{ \partial^j \psi' }]\\
    &= \Re [ \partial_i \alpha \partial_j \alpha  \braket{\psi}{ \psi }+\braket{\partial^i \psi}{ \partial^j \psi } - i \partial_i \alpha \braket{\psi}{\partial_j \psi} + i \partial_j \alpha \braket{\partial_i \psi}{ \psi}] \\
    &= \gamma_{ij}+ \beta_i \partial_j \alpha+ \beta_j \partial_i \alpha + \partial_i \alpha \partial_j \alpha
\end{align}
are not the same after the gauge transformation. 
Here $\beta_i=- i \braket{\psi}{\partial_j \psi}$ is real due to the normalization constraint on $\psi$.
By noticing that $\beta_i$ transforms as
\begin{align}
    \beta'_i &= - i \ \braket{\psi'}{\partial_j \psi'}\\
    &= - i  \braket{\psi}{i (\partial_i \alpha) \psi + \partial_i \psi} \\
    &= \beta_i + \partial_i \alpha~,
\end{align}
and computing
\begin{equation}
    \beta'_i \beta'_j = \beta_i \beta_j + \partial_i \alpha \partial_j \alpha + \beta_i \partial_j \alpha+ \beta_j \partial_i \alpha~,
\end{equation}
we have for \eqref{eq:transormation_metric}
\begin{equation}
    \gamma'_{ij} - \beta'_i \beta'_j =\gamma_{ij} -\beta_i \beta_j := g_{ij}
\end{equation}
Recovering the notation in $\lambda$ we have that $g_{ij}(\lambda)$ is indeed a metric
\begin{equation}
    ds^2 = g^{ij}(\lambda) \dd \lambda_i \dd \lambda_j  = \bra{\delta \psi}(1- \ket{\psi} \bra{\psi})\ket{\delta \psi}~.
\end{equation}
We recall that in general a symplectic structure can be derived from the imaginary part of the scalar product giving the so-called Kähler manifold structure.
\bibliographystyle{unsrt}
\bibliography{bibliography}
\end{document}